\documentclass{ws-procs9x6}

\begin{document}

\title{A Physical Model for the joint evolution of high z QSOs and Spheroids}

\author{G.L. GRANATO}

\address{INAF-Padova, Vicolo osservatorio 5, I-35122 Padova, Italy}

\author{L. SILVA }

\address{INAF-Trieste, Via Tiepolo 11, I-34131 Trieste, Italy}

\author{LUIGI DANESE}
\address{SISSA, Via Beirut 4, I-34014 Trieste, Italy}

\author{G. DEZOTTI}
\address{INAF-Padova, Vicolo Osservatorio 5, I-35122 Padova, Italy}

\author{A. BRESSAN}
\address{INAF-Padova, Vicolo Osservatorio 5, I-35122 Padova, Italy}


\maketitle

\abstracts{We summarize our physical model for the early
co-evolution of spheroidal galaxies and of active nuclei at their
centers. Our predictions are in excellent agreement with a number
of observables which proved to be extremely challenging for all
the current semi-analytic models, including the sub-mm counts and
the corresponding redshift distributions, and the epoch-dependent
K-band luminosity function of spheroidal galaxies. Also, the black
hole mass function and the relationship between the black hole
mass and the velocity dispersion in the galaxy are nicely
reproduced. The mild AGN activity revealed by X-ray observations
of SCUBA sources is in keeping with our scenario, and testify the
build up of SMBH triggered by intense star formation.}

\section{Introduction}

The standard Lambda Cold Dark Matter ($\Lambda$CDM) cosmology is a
well established framework to understand the hierarchical assembly
of dark matter (DM) halos. Indeed, it has been remarkably
successful in matching the observed large-scale structure. However
the complex evolution of the baryonic matter within the potential
wells determined by DM halos is still an open issue, both on
theoretical and on observational grounds.

Full simulations of galaxy formation in a cosmological setting are
far beyond present day computational possibilities. Thus, it is
necessary to introduce at some level rough parametric
prescriptions to deal with the physics of baryons, based on
sometimes debatable assumptions. A class of such models, known as
semi-analytic models, has been extensively compared with the
available information on galaxy populations at various redshifts
(see Granato et al.\ 2000 and references therein).

The general strategy consists in using a subset of observations to
calibrate the many model parameters providing a heuristic
description of baryonic processes we don't properly understand.
Besides encouraging successes, current semi-analytic models have
met critical inconsistencies which seems to be deeply linked to
the standard recipes and assumptions. These problems are in
general related to the properties of elliptical galaxies, such as
the color-magnitude and the [$\alpha$/Fe]-M relations (Cole et al.
2000; Thomas 1999), and the statistics of sub-mm and deep IR
selected (I- and K-band) samples).

These data would be more consistent with the traditional
``monolithic" scenario, according to which elliptical galaxies
formed most of their stars in a single burst, at relatively high
redshifts, and underwent essentially passive evolution thereafter.
However the strict ``monolithic" scheme cannot be fitted in a
consistent model for structure formation from primordial density
perturbations.

However, the general agreement of a broad variety of observational
data with the hierarchical scenario and the fact that the observed
number of luminous high-redshift galaxies, while substantially
higher than predicted by semi-analytic models, is nevertheless
consistent with the number of sufficiently massive dark matter
halos, indicates that we may not need alternative scenarios, but
just some new ingredients.

Previous work by our group (Granato et al.\ 2001; Romano et al.\
2002; Granato et al.\ 2004) suggests that a crucial ingredient is
the mutual feedback between spheroidal galaxies and active nuclei
at their centers. important clues to understand the formation and
evolution of spheroids arise from the now well established
correlation between their stellar mass (or velocity dispersion)
and the mass of the supermassive black hole (SMBH) hosted in their
centers, and responsible for high-z quasar activity.

Granato et al.\ (2004, henceforth GDS04) presented a detailed
physically motivated model for the early co-evolution of the two
components, in the framework of the $\Lambda$CDM cosmology. The
model has been built following suggestions of a scenario
previously explored with a partly empirical approach (Granato et
al.\ 2001).

\begin{figure}[ht]
\centerline{\epsfxsize=2.0in\epsfbox{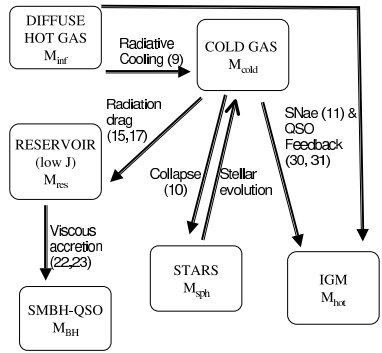}}
\caption{Scheme of the baryonic components included in the model
(boxes), and of the corresponding mass transfer processes
(arrows). The numbers near the arrows point to the main equations
describing those processes in Granato et al.\ (2004).}
\label{fig_schema}%
\end{figure}

\section{Model description}

The model follows with simple, physically grounded, recipes and a
semi-analytic technique the evolution of the baryonic component of
proto-spheroidal galaxies within massive dark matter (DM) halos
forming at the rate predicted by the standard hierarchical
clustering scenario for a $\Lambda$CDM cosmology. The main
difference with respect to other models is the central role
attributed to the mutual feedback between star formation and
growth of a super massive black hole (SMBH) in the galaxy center.
Indeed, the treatment include plausible prescriptions for the
chain of processes leading to SMBH growth trough accretion, and
for the effects on the ISM of the ensuing QSO activity. A scheme
of the baryonic components included in the model, and of the
relevant transfer processes is shown in Fig.\ \ref{fig_schema}.

\begin{figure}[ht]
\centerline{\epsfxsize=2.2in\epsfbox{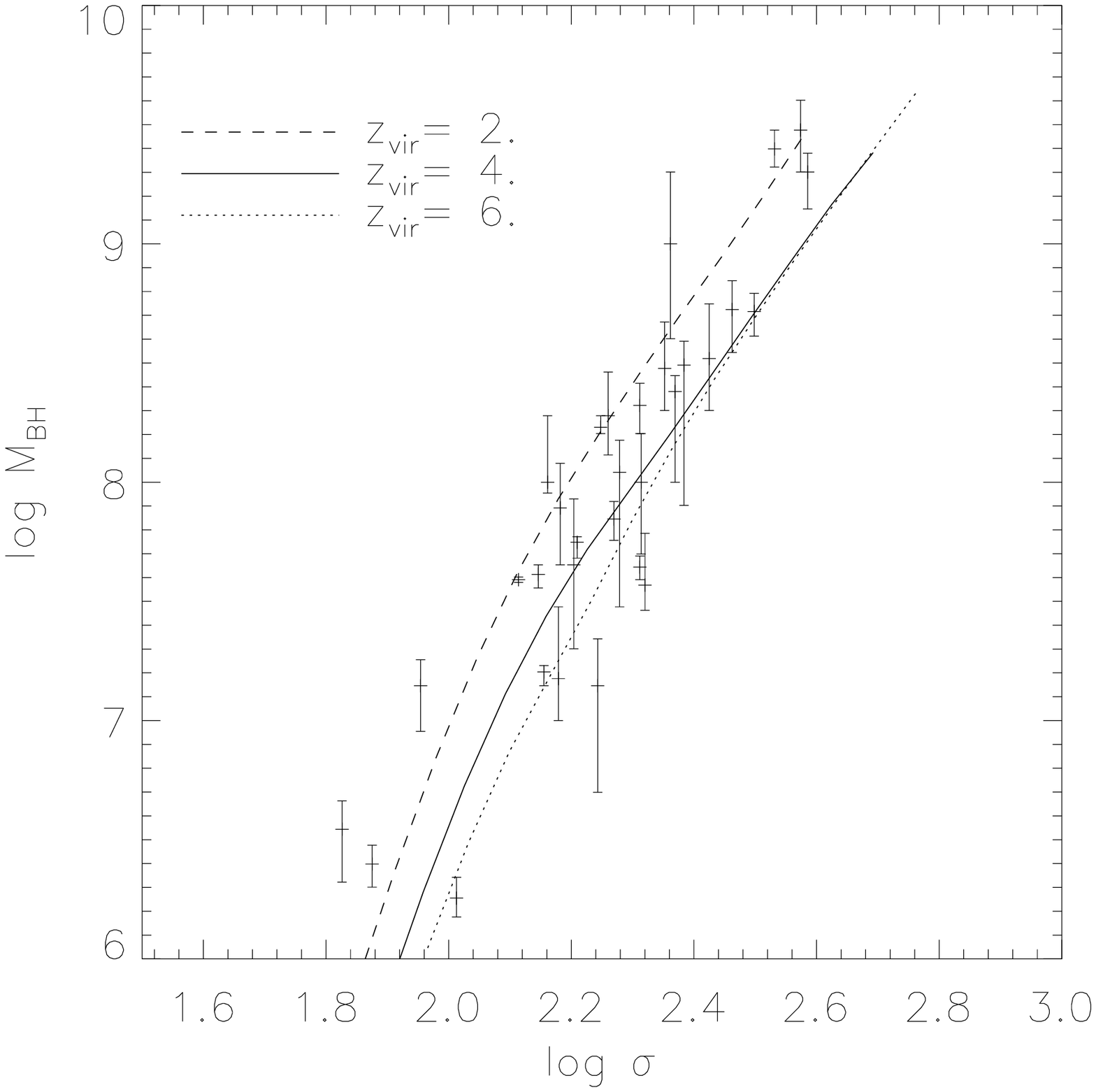}\epsfxsize=3.in\epsfbox{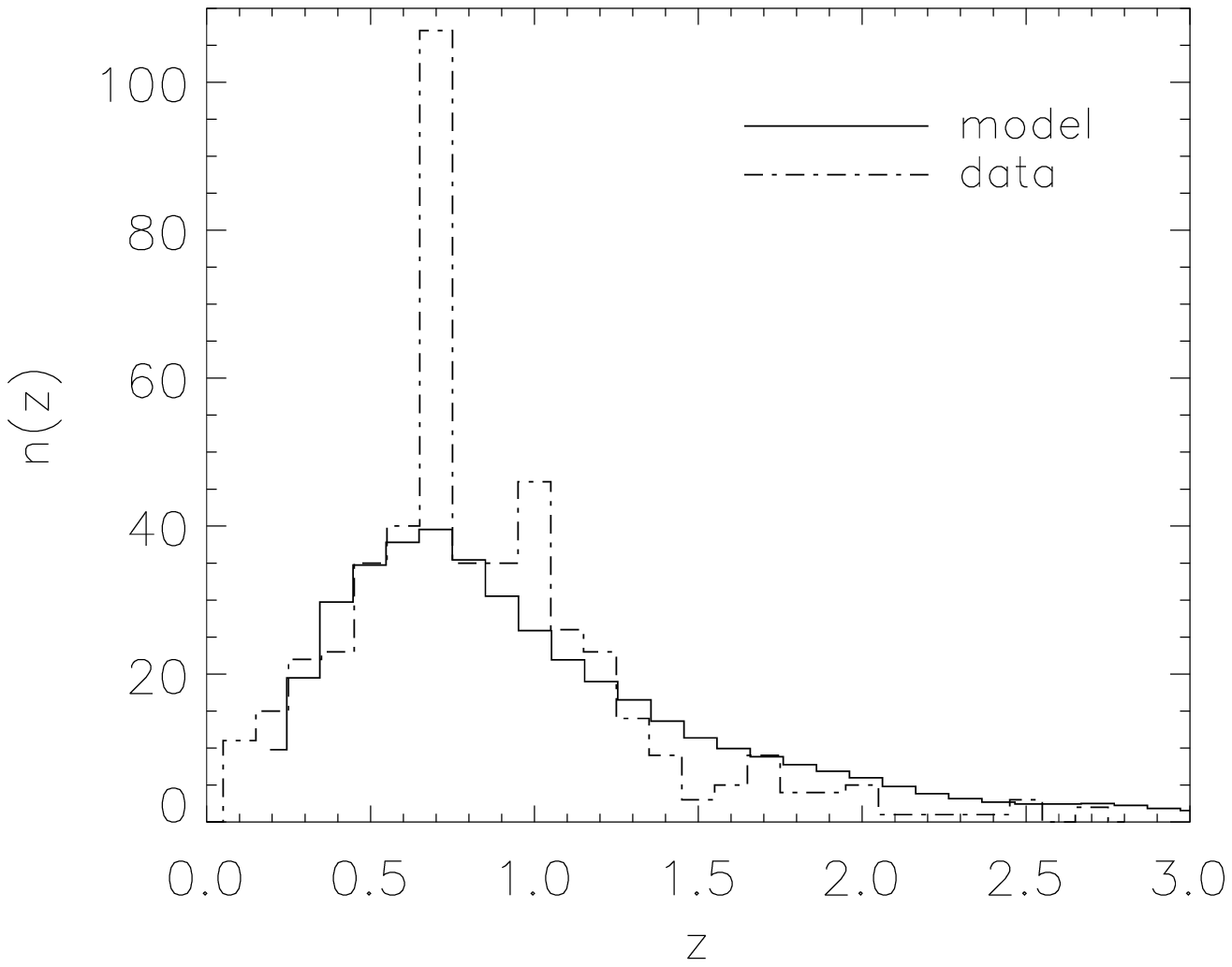}}
\caption{Left: predicted relationship between black-hole mass and
line-of-sight velocity dispersion of the host galaxy for different
virialization redshifts. Right: Predicted redshift distribution of
galax- ies brighter than K = 20 compared with the re- sults of the
K20 survey} \label{fig_mbhsigma_k20}
\end{figure}

The kinetic energy fed by supernovae is increasingly effective,
with decreasing halo mass, in slowing down (and eventually
halting) both the star formation and the gas accretion onto the
central black hole. On the contrary, star formation and black hole
growth proceed very effectively in the more massive halos, until
the energy injected by the active nucleus in the surrounding
interstellar gas unbinds it, thus halting both the star formation
and the black hole growth (and establishing the observed
relationship between black hole mass and stellar velocity
dispersion or halo mass, see Fig.\ \ref{fig_mbhsigma_k20}). As a
result, the physical processes acting on baryons reverse the order
of the formation of spheroidal galaxies with respect to the
hierarchical assembling of DM halos, in keeping with the previous
proposition by Granato et al.\ (2001).

Not only the black hole growth is faster in more massive halos,
but also the feedback of the active nucleus on the interstellar
medium is stronger, to the effect of sweeping out such medium
earlier, thus causing a shorter duration of the active
star-formation phase (for more details, see GDS04).

According to GDS04 (as well as Granato et al.\ 2001), the high
redshift QSO activity marks and concur to the end of the major
episode of star formation in spheroids. Thus there is a clear
evolutionary link between the SCUBA sources and high-z QSOs.
Indeed, the proposed scenario is based on a close and circular
relationship between star formation activity, BH growth and
feedback of the AGN activity on star formation. This relationship
manifest itself as a a well defined and distinctive sequence
connecting various populations of massive galaxies: (i)
virialization of DM halo; (ii) vigorous and rapidly
dust-enshrouded star formation activity, during which a central
SMBH grows; (iii) QSO phase halting subsequent star formation and
(iv) essentially passive evolution of stellar populations, passing
through an Extremely Red Object (ERO) phase. As demonstrated by
GDS04, this scenario fits nicely two very important populations at
high redshift, which are extremely problematic for standard
semi-analytic models (e.g.\ Somerville, 2004): vigorously star-
forming, dust-enshrouded starbursts (in practise SCUBA sources;
stage (ii)) and quiescent red spheroids (stage iv). Also, the
epoch dependent luminosity function of spheroids and the local
mass function of SMBHs is well reproduced.  On the other hand, the
general consistence of this sequence with high redshift QSO
population has been investigated by Granato et al (2001), while a
detailed analysis is the subject of papers in preparation. In the
next section, we analyze as an example the SMBH growth during
stage (ii), as traced by X-ray observations of sub-mm selected
sources.

\begin{figure}[ht]
\centerline{\epsfxsize=3.0in\epsfbox{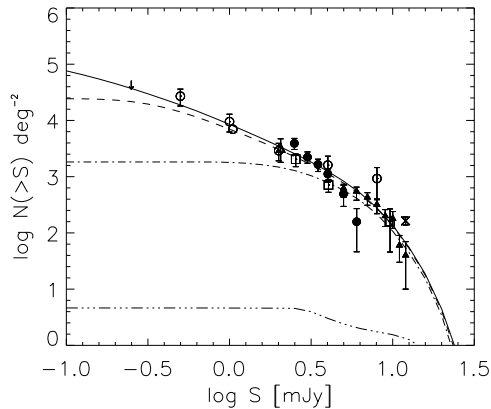}}
\caption{Number counts predicted by GDS04 model for SCUBA sources
with accretion rates greater than several thresholds. Solid: all
sources; dash: $\dot M
> 0.02 M_{\odot} \mbox{yr}^{-1}$; dot-dash: $\dot M > 0.2
M_{\odot} \mbox{yr}^{-1}$; three dot-dash: $\dot M
> 1 M_{\odot} \mbox{yr}^{-1}$ ($L_{\rm bol} > 10^{46} \mbox{erg
s}^{-1}$). See text for explanations.}
\label{fig_counts}%
\end{figure}

\section{AGN sctivity in SCUBA sources}

The model predicts the overall time development of AGN activity in
forming spheroids, though precise predictions in most
electromagnetic bands are made uncertain by environmental effects,
that can significantly influence the way this activity shows up.
This is particularly true in our scenario, since  the QSO growth
occurs in a rather extreme ambient, with no obvious analog in the
local universe. The situation is relatively more favorable with
X-ray photons, especially HX ones, which are the less affected by
interactions with the ISM, and are less likely to be confused with
those produced by processes directly connected with SF, such as
X-ray binaries.

Recently Alexander et al. (2003) noticed that a fraction  $\gtrsim
30-50\%$ of bright ($>5$ mJy) SCUBA sources hosts mild AGN
activity, with X ray (0.5-8 keV) {\it intrinsic} luminosity
between $10^{43}$ and $10^{44}$ erg s$^{-1}$. Using a plausible
bolometric correction of $L_{\rm bol}/L_{X}[0.5-8 {\rm keV}]
\simeq 20$ (Marconi et al.\ 2004) and with the accretion
efficiency 0.1-0.15 adopted by GDS04 (quite standard), these
figures translate into accretion rates onto the central SMBH of
0.02-0.2 $M_{\odot}$ yr$^{-1}$. Fig.\ \ref{fig_counts} shows
number counts for SCUBA sources with accretion rates greater than
several thresholds, and demonstrates that our model is fully
consistent with Alexander et al.\ findings. Indeed, almost all
SCUBA sources brighter than $\simeq$ 5 mJy are expected to host an
AGN with {\it intrinsic} $L_{X}[0.5-8 {\rm keV}] > 10^{43}$
(dashed line in Fig.\ \ref{fig_counts}), leaving room for sources
with high  column density.

According to our interpretation, the moderate AGN activity
revealed by X-ray observations in many bright SCUBA sources
corresponds to the build up by accretion of the central SMBH,
induced by star formation, and well before the bright QSO phase
that cause the end of the major epoch of star formation in these
objects.

\section*{Acknowledgments}
G.L. Granato and L. Silva thank INAOE for financial support and
kind hospitality.

\end{document}